\documentclass{he_symp}
\usepackage{psfig,graphicx,epsfig}
\usepackage{color}
\usepackage{amsmath,amssymb,epic,eepic,array}
\begin{document}
\renewcommand{\FirstPageOfPaper }{ 230}\renewcommand{\LastPageOfPaper }{ 239}

\title{Coherent Radio Emission from Pulsars}
\author{Y.E.Lyubarsky}  
\institute{Ben-Gurion University, P.O.B. 653, Beer-Sheva 84105, Israel}
\maketitle

\begin{abstract}
Generation of the pulsar radio emission from plasma waves excited by
 the two-stream instability is considered. Special attention is given
to propagation effects.
\end{abstract}

\section{Introduction}
Strong coherent radio emissions from pulsars are believed to result from
the development of plasma instabilities in the magnetosphere of the neutron
star. The theory should explain a) what plasma instability excites
collective plasma motions in the pulsar magnetosphere; b) how these collective 
motions generate radio waves and c) how the generated radio waves 
propagate through the magnetosphere. Such a complete theory has not available 
yet; moreover there is still no consensus on the basic emission mechanism.    
The two-stream instability, which
readily excites strong plasma oscillations, is among the most widely discussed.
 Pulsar magnetospheres are believed to be 
filled with an electron-positron plasma generated in a cascade process from
a primary particle beam accelerated in the rotationally induced electric field
(see the recent work by Hibschman \& Arons 2001 and references therein). 
This plasma streams along the open magnetic field lines, which extend beyond the light cylinder.
Oscillations generated in this plasma may be a source of a powerful
radio emission.  

Conditions for the development of the two-stream instability in pulsar
magnetospheres are not trivial. The plasma flow should involve
 streams with markedly different velocities. Usov (1987)
pointed out that such a
 configuration arises if plasma production is extremely unsteady. Then the plasma
flow is built up from separate clouds following each other along the
magnetic field line and slower particles from a cloud would be overtaken by
faster particles from the subsequent cloud giving rise to the two-stream
instability (see also Ursov \& Usov 1988; Asseo \& Melikidze 1998).
A two-stream configuration arises also because the adjustment of electric 
current and charge density in the open field line tube requires a certain 
flow of particles to be directed downwards, against the main plasma flow 
(Lyubarskii 1992a,b; 1993a). At last conditions for the two-stream 
instability may arise as a result of interaction of the secondary
pair plasma with the thermal emission from the neutron star surface (Lyubarskii
\& Petrova 2001). Discussion of these scenarios is out of the scope of 
my talk; it is enough to mention that in the plasma flow within the open 
magnetic line tube, the two-stream instability develops readily. Therefore
let us assume that the instability does develop and consider
 the formation of pulsar radio emission from the excited plasma waves and the
properties of the outgoing radiation.

\section{Generation of radio emission from plasma oscillations in pulsar 
magnetospheres}
\subsection{Waves in the plasma embedded in a super strong magnetic field}
Let us first outline qualitative properties of waves in the
electron-positron plasma embedded in a super strong magnetic field.
The pulsar radio beam is believed to
originate from deep inside the magnetosphere, where the magnetic field 
may be considered as effectively infinite. Namely particles in such a 
field may be considered as beads on a wire; they move freely
along the magnetic field lines and do not shift in the transverse direction. 
Properties of this plasma are relatively simple. They are especially
simple if plasma is cold therefore below all estimates will be made 
in this limit. In the relativistically hot
plasma properties of the waves are qualitatively the same (see, 
e.g., Volokitin et al.\ 1985; Arons \& Barnard 1986; Lyubarskii 1996;
Melrose \& Gedalin 1999; Asseo \& Riazuelo 2000).
 
Let us first consider waves in the plasma rest frame
(all the quantities in this frame will be marked by prime). It is evident that
a transverse wave propagating along the magnetic field does not interact with
the plasma because the electric field in this wave is perpendicular
to the background magnetic field and  plasma particles are unable to
move in this direction. Therefore transverse waves propagate along a
super strong magnetic field just like in vacuum; for these waves $\omega=kc$.
On another hand, there exist longitudinal, electrostatic waves propagating 
along the magnetic
field and these waves are just the same as Langmuir waves in a 
nonmagnetized plasma because
particles oscillating along the magnetic field do not "feel" this field. So
 frequency of such a wave is equal to the plasma frequency (in the plasma
rest frame) independently of the wavelength, 
$$
\omega_p=\sqrt{\frac{4\pi e^2n'}{m}},
\eqno(1)
$$
 just as in a nonmagnetized plasma. Here $n'$ is the plasma number density
in the proper frame, $e$ and $m$ the electron charge and mass, respectively.

\begin{figure}
\centerline{\psfig{file=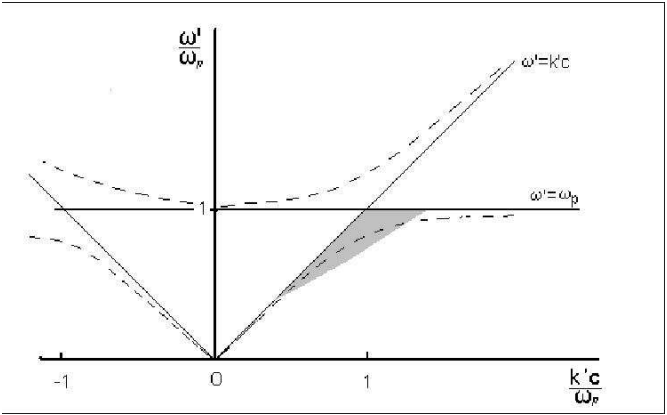,width=8.8cm,clip=} }
\caption{Dispersion curves in the proper plasma frame. Solid lines correspond
to the waves propagating along the magnetic field and dashed, to those
propagated obliquely. The two-stream instability excites waves in the shaded
region.
\label{image}}
\end{figure}
\begin{figure}
\centerline{\psfig{file=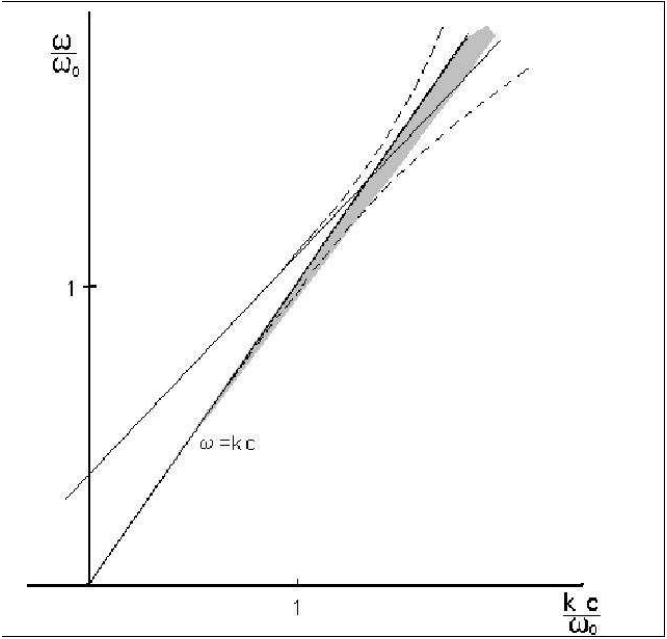,width=8.8cm,clip=} }
\caption{Dispersion curves in the pulsar frame. 
\label{image}}
\end{figure}

 The dispersion curves of the waves propagating along the magnetic
field are shown in Fig.\ 1 by solid lines. The horizontal line corresponds to the
longitudinal plasma wave and the bisector corresponds to the transverse
wave. 
Intersection of the dispersion curves (degeneracy) is possible because
the system is axisymmetrical for the waves propagating along the magnetic field.
The dispersion curves of the oblique waves can not intersect because axisymmetry 
is broken in this case. Taking into account that 
properties of waves propagating at small angles to the magnetic 
field should be close, by continuity, to properties of waves 
propagating strictly along the magnetic field, one can sketch the dispersion
curves of oblique waves (dashed lines in Fig.1). Of course  
 this simple qualitative picture is confirmed by calculations (see the 
above-mentioned papers).
One can see that the dispersion curves of the oblique
waves are split into  superluminous ($\omega'>k'c$) and subluminous
($\omega'<k'c$) branches, the separation between the branches being larger for
waves propagating at larger angle to the magnetic field. Polarization
of  oblique waves is neither purely longitudinal nor transverse however 
polarization of  waves propagating at small angle to the magnetic field
is close to longitudinal or transverse in the corresponding parts of the
dispersion curve. At $\omega'\gg\omega_p$ the superluminous wave becomes
nearly transverse wave with the vacuum dispersion law, $\omega=kc$.
It is very important that a single branch of the dispersion curve for the
oblique wave links nearly longitudinal and nearly transverse waves.

In pulsars plasma streams with a Lorentz factor $\gamma\sim 100$ along the
open magnetic field
lines. The dispersion curves in the pulsar frame may be obtained by Lorentz
transformation of the curves shown in Fig.\ 1. The result is plotted in Fig.\ 2.
The dispersion law of the transverse wave propagating along the magnetic field, $\omega=kc$, evidently remains  unchanged whereas the horizontal line
$\omega'=\omega_p$, corresponding to the longitudinal waves, transforms into
an inclined line such that the frequencies of the longitudinal plasma waves may 
vary significantly. The reason is that not only the frequency but also the
  wavelength
enter the Lorentz transformation therefore the waves with different
wavelengths and the same frequency in the proper plasma frame have different
frequencies in the pulsar frame. 
The wave $\omega'=\omega_p$, $k'=0$ have the frequency
$$
\omega_0=\omega_p\sqrt{\gamma}
\eqno(2)
$$
 in the pulsar frame. The waves with $\omega<\omega_0$ are those propagating
in the proper frame upstream the plasma flow. 
The dispersion curves of the transverse and longitudinal waves propagating
along the magnetic field intersect at the point $\omega=2\omega_0$, 
$k=2\omega_0/c$; above this point the longitudinal wave is subluminous.
 The dispersion curves for the oblique waves are also 
split into superluminous and subluminous branches. At $\omega\gg \omega_0$
the superluminous wave becomes the vacuum transverse electromagnetic wave.

The waves resonantly interact with particles if the phase velocity, $\omega/k$, 
coincides with the particle velocity (more exactly the resonance condition
reads as $\omega={\bf k\cdot v}$). Of course only subluminous waves
participate in such an interaction. In the presence of a particle beam, such a 
resonant interaction leads to
the two-stream instability, which predominantly excites waves with the
phase velocities close to the beam velocity. The region where waves are
generated by a high velocity particle beam is shaded  in Fig.\ 1. Waves with low
enough phase velocities are heavily damped because they are in resonance with 
the thermal plasma particles (Landau damping). 

Up to now only so called ordinary waves were considered. There is also
an extraordinary wave; in this wave the electric field is perpendicular both
to the background magnetic field and to the wave vector $\bf k$. In
the limit of infinitely strong magnetic field this wave evidently does not
interact with the plasma and propagates like in vacuum.

\subsection{Escape of waves}
The crucial question is what waves can escape from the pulsar magnetosphere
and form the observed radio beam. The answer is 
evident for the nonmagnetized plasma where the purely
longitudinal as well as purely transverse waves exist.  
Namely the longitudinal wave, which is electrostatic, do not escape
whereas the transverse wave escapes freely. There is no purely transverse 
waves in the strongly magnetized plasma and therefore one should ask what
wave becomes the vacuum transverse electromagnetic wave when it  
propagates outward, in the plasma of decreasing density (Arons \& Barnard 1986). 

Let us consider propagation of a superluminous wave.
In a steady state medium with smooth density gradients, the wave frequency
remains constant while the wavelength is adjusted in order to satisfy the dispersion
equation at any point. In the $(\omega, k)$ plane, the point representing a
wave moves along the dispersion curve upward because plasma frequency decreases
together with plasma density. If the wave propagates strictly along the magnetic
field, it transits along the dispersion curve to the subluminous region. Eventually
the wave phase velocity, $\omega/k$, decreases such that the wave
 decays  through the Landau damping. In the
oblique propagation case, the wave remains superluminous and at
$\omega\gg\omega_0$ transforms into the vacuum transverse electromagnetic wave,
which escapes freely \footnote{Of course
an infinitely small angle between the magnetic field
and the propagation direction is insufficient for free escape of a
longitudinal wave. If the angle is small enough,
the superluminous branch approaches closely to the subluminous one near the point
$\omega=2\omega_0,\quad k=2\omega_0/c$ and the wave may jump, because of linear coupling, onto the 
subluminous branch.
However in pulsar magnetospheres, this effect is of no importance
because the corresponding critical angle is very small
(Bliokh \& Lyubarskii 1996).}.
Finally it should be noted that  in a curved magnetic field, a wave becomes
oblique, even if it was initially directed along the
magnetic field (Barnard \& Arons 1986, Lyubarskii \& Petrova 1998). Therefore 
the superluminous longitudinal
waves eventually escape from pulsar magnetospheres in the form of the vacuum
 transverse electromagnetic waves.

Applying the same consideration to the subluminous waves, one can  see
that such waves do not escape.  When they propagate in the plasma of
decreasing density, their  phase velocities decrease and eventually 
the waves decay through the Landau 
damping. So in the strong magnetic field the superluminous wave escapes even 
if it was initially purely longitudinal (in the curved magnetic field)
whereas the subluminous waves does not escape even though these waves are nearly
transverse at $\omega\ll\omega_0$.

The two-stream instability generates predominantly longitudinal waves, which 
are in Cherenkov resonance with the particle beam, $\omega=kv$. This waves 
are evidently
subluminous, $\omega/k<c$ (the region where the waves are generated is shaded
in Figs.\ 1, 2).   Such waves are unable to escape from plasma unless nonlinear 
processes redistribute the wave energy into the superluminous region. High
brightness temperature of pulsar radio emission implies high wave energy 
density in pulsar magnetospheres and therefore nonlinear effects should be of
paramount importance.

\subsection{The second order processes}
As the first step in studying of the nonlinear effects, one should
consider the processes of the lowest order in the wave 
energy density, $W$. These are the wave-wave interaction and induced scattering
by the plasma particles (see, e.g., Tsytovich 1970; Melrose 1980). The 
wave-wave 
interactions involve, in the lowest order in $W$, merging of two waves into one
and decay of a wave into two waves.  All three interacting waves should
satisfy the dispersion equation, $\omega=\omega({\bf k})$, and also the energy and 
momentum conservation laws,
$$
\omega({\bf k}_1)+\omega({\bf k}_2)=\omega({\bf k}_3);
$$
$$
{\bf k}_1+{\bf k}_2={\bf k}_3;
$$
which place severe restrictions on the process. Therefore this process is
rather ineffective in pulsar conditions (Bliokh \& Lyubarskii 1997)
\footnote{ There are more degrees of freedom for wave-wave interactions in a 
finite magnetic
field (Machabeli 1983; Lyutikov 1999) however probabilities of these processes 
are rather small.} . 

In contrast to wave-wave interactions, there is no restrictions on induced scattering and therefore
this process is of crucial importance. 
At rather general conditions, the induced scattering redistributes the wave energy towards large
wavelengths (in the plasma rest frame). One can easily see from Fig.\ 1 that
when a longitudinal wave from the shaded region moves towards 
small $k$ (larger
wavelengths) it eventually becomes superluminous. However it was shown above
that superluminous waves  escape freely in the form of vacuum transverse
waves if they propagate in the curved magnetic field in the plasma of
decreasing density. So the observed radio emission may be 
generated through
the induced scattering of the longitudinal waves excited by the two-stream
instability.

Redistribution of the wave energy is described by the kinetic equation;
in the  plasma in the infinite magnetic field it may be presented in the form
$$
{{\partial W_{\bf k}}\over{\partial t}}+
{\bf v}_g {{\partial W_{\bf k}}\over{\partial {\bf r}}} 
\eqno(3)
$$$$
=
W_{\bf k}\int w_p ({\bf k,k_1}) W_{\bf k_1}
{{k_z -k_{z1}}\over{\omega_1}}{{\partial f}\over{\partial p}} dpd{\bf k}_1,
$$
where $W_{\bf k}$ is the spectral energy density of waves, $f(p)$ the particle
distribution function,
$w_p ({\bf k,k_1})$ the  probability of the scattering of the wave $\bf k_1$
into the wave $\bf k$, ${\bf v}_g$ the group velocity of the wave; the field is
directed along $z$ axis.  The scattering probability was calculated by
Lominadze et.\ al (1979) for waves propagating along the magnetic field
and by Lyubarskii (1993b) in the general case.

One can conveniently analyze the plasma processes in the proper plasma frame.
In this frame
(more exactly, in the frame of reference moving with the maximum of the
particle distribution function) ${{\partial f}\over{\partial p}}\le 0$ and
the right-hand side of Eq.(3) is positive if $k_z<k_{k1}$, so
the wavelength increases  in the scattering process. 
 Kinetics of the 
induced scattering in the one-dimensional relativistic plasma was considered by 
Lyubarskii (1996). Of course in the presence
of a particle beam, one more maximum appears in the distribution function and, 
provided
the beam is strong enough, the above statement violates and the wave frequency 
may significantly increase (e.g., Weatherall 2001, Schopper et al.\ 2002). 
In this case the beam
plays the role of the main plasma and, switching to the frame of the beam,
we come to the same conclusion: the wavelength increases. 
In the frame of the beam this is accompanied by changing  of the wave
propagation direction and therefore the wave frequency grows in the laboratory
frame. One should also note that the 
so called amplified linear acceleration emission (Melrose
1978; Rowe 1995) actually represents the induced scattering of a longitudinal,
electrostatic wave into a (quasi) transverse one.  

For  a rough estimate, one can present
the kinetic equation (3) in the simple form
$$
{{dW}\over{dt}}\sim \eta W,
\eqno(4)
$$
where $W=\int W_{\bf k}d{\bf k}$ is the total wave energy density. In the
proper plasma frame, the redistribution rate  may be roughly 
presented as
$$
\eta\sim\frac{W'}{n'mc^2}\, \omega_p.
\eqno(5)
$$
The structure of this simple estimate is common for any second order process:
the corresponding redistribution rate is proportional to the ratio of the
wave energy to the plasma energy density and some relevant frequency. In 
the case of interest, the characteristic plasma "temperature" is about of
the electron rest energy and all frequencies are about the plasma frequency
therefore the estimate is especially simple and may be in fact obtained from
dimensional considerations.

\subsection{Emission power}
The mechanism under consideration generates electromagnetic waves with 
 $k'<\omega_p/c$  (see Fig.\ 1); in the pulsar frame the frequencies
of these wave are about $\omega_0$ (Eq.(2)). So the observed frequency 
is determined,
via the plasma frequency, by the plasma density at the emission point. The 
electron-positron plasma is assumed to be generated at the base of the 
open field line tube; the generation rate is conveniently characterized by
the multiplicity parameter, $\kappa$, such that the plasma density at the
base of the open field line tube is normalized by the Goldreich-Julian (1969)
density,
$$
n=\kappa\frac{B}{ecP},
\eqno(6)
$$
where $B$ is the surface magnetic field, $P$ the pulsar period. Continuity
of the plasma flow in the tube implies that the plasma density
falls off as the tube cross section; for the dipole field, $n\propto 1/r^3$.
Now the radiation frequency may be estimated as
$$
\nu=\frac{\omega_0}{2\pi}=1.4\sqrt{\frac{\kappa_2\gamma_2B_{*12}}{P}
\left(\frac{30r_*}{r}\right)^3}\, {\rm GHz},
\eqno(7)
$$
where 
$B_{*}\equiv 10^{12}B_{*12}$ G is the surface magnetic field, $r_*$ the
stellar radius, $\kappa_2\equiv\kappa/10^2$, $\gamma_2\equiv\gamma/10^2$.
Observational data suggest that the radio emission is generated at  altitudes 
about few hundreds km
(e.g., von Hoensbroech  \& Xilouris  1997; Kijak \& Gil 1998; Kijak 2001). At
such altitudes,  the width of the open field line 
tube is about the observed beam width, the last being estimated as
 $\vartheta=0.1P^{-1/2}$ (Rankin 1993). The estimate (7) shows that the radio
emission may be generated at the plasma frequency
provided the multiplicity is about 100. Note that the presented 
radius-to-frequency mapping does not mean that the emission is narrow band.
Plasma at a given radius radiates in the range $\Delta\nu\sim\nu$.

Because nonlinear redistribution rates depend on the wave energy density,
one can easily estimate the radiation power comparing the redistribution 
rate with the escape rate. The two-stream instability generates Langmuir
waves in the region shaded in Fig.\ 2. These waves propagate upward in the 
 plasma of decreasing  density; without nonlinear 
redistribution, the phase velocity of the waves would decrease and the waves 
would eventually decay. Phase velocity decreases with plasma density at
the characteristic scale $r$; in the proper plasma frame, the corresponding 
time is about $r/(c\gamma)$. Emission of radio waves is possible if the
induced scattering transforms the waves into superluminous ones for a lesser
time, $\eta \ge r/(c\gamma)$. Making use of Eq.\ (5), one can estimate
 the minimal wave energy density necessary for the emission  as
$$
\frac{W'}{n'mc^2}=\frac{W}{\gamma nmc^2}=
0.016\frac{\gamma_2^{5/3}P^{1/3}}{(\nu_9\kappa_2B_{*12})^{1/3}}.
\eqno(8)
$$
Here the radius-to-frequency mapping (7) was used to eliminate the radius in
favor of the radiation frequency. If the plasma wave density is less then the above
value, nonlinear processes are ineffective and the waves eventually decay. The 
pulsar radiates if the two-stream instability generates waves with the energy 
density exceeding the minimal value; then the generated waves are transformed into
superluminous ones and may escape.  The corresponding luminosity may be
estimated multiplying the wave energy density  by the speed of light and the
cross section of the polar tube, $S=\pi r_*^3\Omega/c$; for the two tubes one gets
$$
L_{min}=2WcS=5\cdot10^{26}\frac{(\kappa_2B_{*12})^{2/3}\gamma_2^{8/3}}
{P^{5/3}\nu_9^{1/3}}\,{\rm erg/s}.
\eqno(9)
$$
The presented estimate roughly corresponds to the luminosity of weak pulsars. The
radiation spectrum in this case may be estimated as 
$\frac{dL}{d\nu}\propto \nu^{-4/3}$.

The pulsar luminosity is limited from above by the total energy of the
plasma flow. So the maximal pulsar luminosity may be estimated as
$$
L_{max}=2nmc^3\gamma S=2.5\cdot 10^{28}\frac{\kappa_2B_{*12}\gamma_2}{P^2}\,
{\rm erg/s}.
\eqno(10)
$$
This estimate is compatible with the observed pulsar luminosities. 

According to Eq.(7), the plasma multiplicity factor should be not large, 
$\kappa\le 100$, for radio emission to be emitted not too high in the 
magnetosphere.
Originally the plasma production models predicted $\kappa\sim 10^3-10^4$ and
even more (Ruderman \& Sutherland 1975; Daugherty \& Harding 1982; Arons 1983).
However recent calculations, which take into account the resonant scattering of
the thermal radiation from the surface of the star, favor for smaller $\kappa$
(Hibschman \& Arons 2001), which are in better agreement with the assumption
that the emission is generated at the plasma frequency.
On the other hand, the observed pulsar luminosities require, according
to Eq.(10), $\kappa\ge 100$. Whilst these estimates are not incompatible,
one would nevertheless fill more comfortable if the plasma could emit at 
frequencies well below the plasma frequency or if there was an additional 
energy source such
that a low density plasma could provide the observed radiation power. 

Radiation at
$\nu\ll\nu_p$ is possible, in principle, in the subluminous mode 
(Arons \& Barnard 1986; Lyutikov 2000; Melikidze et al.\ 2000). This wave
propagates along the magnetic field preserving original direction of 
the wave vector. Being emitted  at a  level where the plasma density is high,
the wave is ducted upwards along the magnetic field line and eventually decays
when the phase velocity reaches the particle velocities (Barnard \& Arons 1986).
The outgoing radiation may be generated if the wave is transformed into  
the superluminous wave before the decay takes place.  One should expect that 
then the beam width will be determined by the width of the tube at 
the transformation level, which is in fact the level where $\nu\sim\nu_p$.
So if $\kappa$ is large, the beam should be too wide even if it was emitted
at low altitudes. The beam will remain 
narrow only if  the transformation process
does not affect the original direction of $\bf k$ however up to now nobody has
proposed such a transformation process.

The possibility that the plasma in the open field line tube gains energy from 
an external source may not be ruled out because   
the  electromagnetic energy in the pulsar magnetosphere 
greatly exceeds the plasma energy. 
True, plasma with $\kappa >1$ screens electric fields and therefore 
the huge electric 
potential generated in the open field line tube seems to remain beyond the 
reach of the 
plasma flow. Nevertheless adjustment of the electric charge density in the open
field line tube requires some longitudinal electric field (Scharlemann 1974;
Cheng \& Ruderman 1977). It is possible that the plasma flow 
may gain some energy from this field; then the pulsar luminosity may exceed
that of Eq.(10).

\subsection{Strong turbulence}
If the wave energy density is about Eq.(8), the induced scattering rate is
comparable with the propagation time and therefore the waves are transformed into 
superluminous ones and escape. If the wave energy is larger, the induced scattering
rate exceeds the escape time, which means that the effective optical depth becomes
large. In this case the waves may be locked in the system therefore another
nonlinear processes come into play.

The most important effect is the modulation instability. A spatially uniform 
distribution of plasma waves becomes unstable with respect to formation of regions
with enhanced energy density of the waves (e.g., ter Haar \& Tsitovich 1981;
Goldman 1984).
In the nonrelativistic plasma, this instability
 develops already at rather small plasma wave energy densities. 
The threshold of the instability is
$$
\frac W{W_{pl}}\sim (kD)^2,
\eqno(11)
$$
where $W_{pl}$ is the plasma energy density, $D$ the Debye length. The waves
are excited by a particle beam with the velocity $v_b$ at the resonance 
$k=v_b/\omega_p$ therefore the right hand side of Eq.\ (11) may be presented as
$(v_{th}/v_b)^2$, where $v_{th}$ is the electron thermal velocity. Taking into account that
typically $v_b\gg v_{th}$, one can see that the threshold of the instability
is rather low.
 In the relativistic plasma, $v_b\sim v_{th}\sim c$
and $W'_{pl}\sim n'mc^2$ therefore the modulation instability develops only
when the wave energy density becomes comparable with the plasma energy density.
In this case the induced scattering rate (5) is comparable with the plasma frequency
and therefore can not be neglected. Interplay between the two effects should be 
considered, which has not been done yet. One can only anticipate that the induced 
scattering makes the development of the
modulation instability easier because the wavelength increases. 
 
It is still unclear how formation of cavities with the enhanced wave energy density
influences the emission properties of the pulsar plasma. Evidently the
effect results in a
short term variability. However it is still unclear how outgoing waves are generated.
Asseo et al.\ (1990), Asseo (1993), Melikidze et al.\ (2000) considered
emission from the cavities, or solitons, assuming that they are stable.
However Weatherall (1997, 1998)
found numerically that the longwave Langmuir solitons collapse like in the 
nonmagnetized plasma. He suggested that the waves escape from the
collapsing cavities because longitudinal electrostatic waves evolve into
oblique electromagnetic waves in the course of the collapse. However these waves
may be locked within the cavity surrounded by a higher density plasma. 
 More careful analysis is necessary, both
numerical  and analytical, to clarify
the situation. Comparing
analytical estimates with the results of numerical simulations, one will be able
to find reliable scalings, which may be applied to real pulsars.

\section{Propagation effects}
\subsection{Polarization limiting radius}
 If the radio emission is generated by the two-stream instability, the wave
frequency should be of the order of the plasma frequency at the emission point.
Then many observed characteristics of the outgoing 
radiation, first of all polarization,  are dictated by the 
wave propagation in the magnetospheric plasma. Whatever the emission mechanism, 
radiation propagates in the plasma in the form of two orthogonally polarized
normal waves. Deep inside the magnetosphere, normal waves are linearly 
polarized; the ordinary wave is polarized in the plane defined by the 
ambient magnetic field, $\bf B$, and the wave vector, $\bf k$, and the 
extraordinary wave in the $\bf k\times B$ direction. The polarization pattern
follows, along the ray, the local orientation of the $\bf k\times B$
plane; Cheng \& Ruderman (1979) used the term "adiabatic walking" to describe 
the evolution of the ray polarization in this regime. Ultimately the plasma
density falls to an extent that the medium no longer affects wave propagation.
The observed polarization is fixed in the transition region, at the so called
polarization-limiting radius, which is determined by the condition
$$
kL\Delta n = 1,
\eqno(12)
$$ 
where $L$ is the characteristic scale length for changing polarization of the
normal waves, $\Delta n$ the difference in index of refraction between the
normal waves. Because all characteristic scales in the pulsar magnetosphere
are large as compared with the wavelength, $kL\gg 1$,  $\Delta n$ should be
small at the polarization limiting radius, which typically means that 
both indexes are close to unity, i.e.\ the local plasma frequency is much
lower then the wave frequency. So the polarization-limiting altitude should be
significantly higher then the emission altitude.

In pulsar magnetospheres, the polarization-limiting radius was estimated by
Cheng \& Ruderman (1979), Melrose (1979), Barnard (1986). Unfortunately
this value depends not only on the plasma density, which itself
is rather uncertain, but also on the angle the wave vector makes with the
magnetic field. Since this angle is small, even small deviation of the
magnetosphere structure from the pure dipole is of crucial importance. 
Therefore one cannot firmly fix the polarization-limiting radius until
the self-consistent model for the relativistic, rotating magnetosphere will
be available. 

Observations of the polarization position angle swing, i.e.\ of the smooth 
rotation of the plane of linear polarization through the pulse, place limits 
on the polarization limiting radius. The observed 
polarization position angle is determined by the projection of the magnetic
field on the plane perpendicular to the ray direction at the polarization 
limiting radius. Deviations of the magnetosphere structure from the pure dipole
at the polarization-limiting radius, as well as the aberration and retardation 
effects, should affect the position angle vs.\ pulse longitude curve.
The success of the simple rotating vector 
model in description of the position angle swing (Radhakrishnan \& Cooke 1969;
 Manchester \& 
Tailor 1977) suggests that the polarization-limiting radius is small as
compared with the light cylinder radius. In particular, the position angle sweep
decreases with the altitude; this was demonstrated by Barnard (1986) for the Deutch
magnetosphere but this effect is of general nature (Fig.\ 3).
 Large position angle sweeps, which are not uncommon in pulsars, favor 
for a small
polarization limiting radius.  Large sweeps may be obtained at large
altitudes only if 
the magnetosphere sweepback caused by rotation is exactly compensated by
the rotation itself, so the plasma motion is strictly radial and the magnetic 
field in the proper plasma frame is also radial. In this case the
 magnetic axis  has exactly the form of the Archemedean spiral. 
Since the self-consistent model
of the pulsar magnetosphere has not available yet, such a possibility cannot
be excluded however below the general picture sketched in Fig.\ 3 is
assumed.

\begin{figure*}
\centerline{\psfig{file=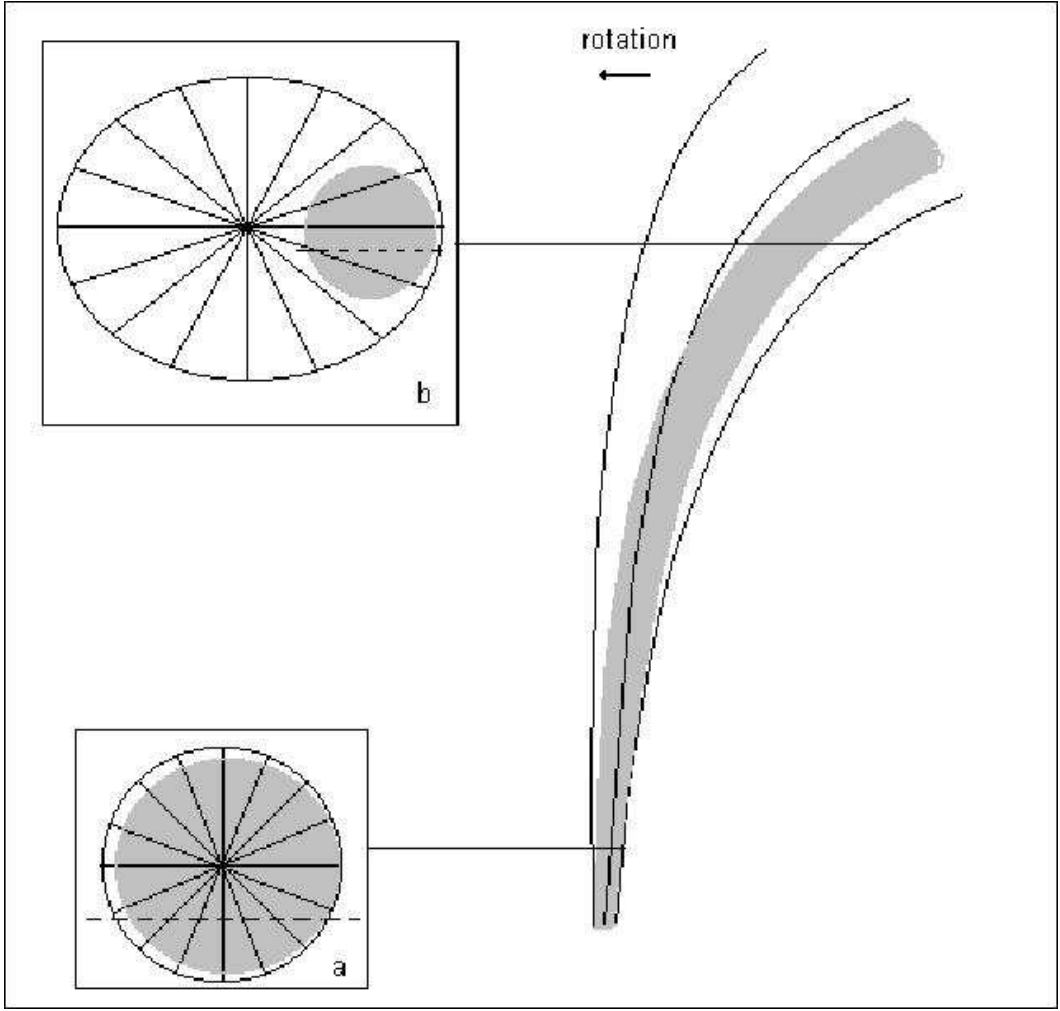,width=14cm,clip=} }
\caption{Propagation of the radio beam within the magnetosphere. The open
field line tube is swept back because of rotation. The shaded region traces
the beam path within the rotating magnetosphere. In the cross sections {\it a}
and {\it b}, solid lines show the magnetic planes, the dashed lines trace the
path of the line of sight in the course of the magnetosphere rotation. One 
can see that in the low altitude cross section {\it a}, the sweep of the
magnetic planes, which determines the polarization position angle sweep, may
be large whereas in the high altitude cross section {\it b} the
sweep is always small.
\label{image}}
\end{figure*}

The position angle sweep decreases with increasing ratio of the 
polarization-limiting radius to the light cylinder radius. Therefore the sweep 
should decrease, in average, with decreasing pulsar period (Barnard 1986). 
This conclusion agrees with shallow polarization swing observed in 
millisecond pulsars (Xilouris et al.\ 1998).

\subsection{Circular polarization}
The radiation from many pulsars contains detectable circular polarization,
although the amounts are generally much less than the degree of linear 
polarization (e.g., Radhakrishnan \& Rankin 1990; Han et al.\ 1998).  The
circular polarization naturally arises at high altitudes where the cyclotron
frequency is already not too large as compared with the radiation frequency. The
normal waves are elliptically polarized in this region if the distribution
functions of electrons and positrons are different. This is always the
case because the Goldraich-Julian charge density should be maintained
in pulsar magnetospheres. Therefore the circular polarization in the outgoing
radiation may be attributed to the dispersive properties of the magnetospheric
plasma provided the polarization-limiting radius is about the light cylinder
radius (Melrose \& Stoneham 1977; Cheng \& Ruderman 1979;
Melrose 1979; von Hoensbroech et al.\ 1998; von Hoensbroech \& Lesch 1999;
Gedalin et al.\ 2001). However large position angle sweeps imply
a small polarization limiting radius, at least in significant fraction of
pulsars. At such radii, the normal waves are linearly polarized and  elliptical 
polarization may arise from the wave mode coupling in the polarization-limiting 
region (Cheng \& Ruderman 1979).

Deep inside the polarization-limiting radius, where the plasma density is high
 enough, the normal waves propagate
independently and their polarization plane is adjusted to the local
orientation of the $\bf k\times B$ plane. Outside this region, the plasma 
density is too low and the waves propagate as in vacuum. In the polarization
limiting region, the normal waves are still influenced by the plasma but
this influence is insufficient to make the wave polarization follow the 
orientation of the $\bf k\times B$ plane. Provided   the 
$\bf k\times B$ plane turns along the ray path, wave mode coupling takes place
resulting in the elliptical polarization of the outgoing waves. The  
$\bf k\times B$ plane turns along the ray path
due to rotation of the magnetosphere, or due to sweepback of the magnetic field
lines caused by the rotation or if the initial co-planarity between $\bf k$ and
the field line fails because of refraction.

The quantitative treatment of the polarization transfer in the rotating
magnetosphere was held by Lyubarskii \& Petrova (1999) and Petrova \&
Lyubarskii (2000). They found significant circular polarization of the 
outgoing radiation, in some cases with the sense reversal near the pulse
center. The degree of circular polarization depends on the limiting
polarization radius and reaches large values, $V\sim 1$, if this radius is
about $r_L\vartheta$, where $\vartheta$ is the beam width, $r_L$ the light 
cylinder radius. 

 Circular polarization is generally larger in the central part of the
pulse profile, in the core beam according to Rankin's (1983a) classification.
This property may be naturally explained if the polarization limiting radius
is small, $<r_L\vartheta$, such that the magnetic axis remains within the beam
(see Fig.\ 3).  In this case the angle between the ray and the magnetic
field decreases towards the center of the beam. The lesser this angle,
the larger variation in the orientation of the  $\bf k\times B$ plane may be
caused by a small deviation from co-planarity between $\bf k$ and the field line
(Fig.\ 4).  Therefore the circular polarization  increases towards the center
of the beam. Moreover violation of the regular position angle swing, which is
commonly observed in the core beams (e.g., Rankin 1990), may be naturally explained 
by the same reason. One should also note that
 the angle between the magnetic field and the ray direction
in the cone beam is also not large, therefore the difference between the 
polarization characteristics of the cone and core beams is not qualitative 
but only quantitative.
This also explains why there is no abrupt transition from
the cone to the core emission but merely a gradation of properties across the
whole emission beam (Lyne \& Manchester 1988; Han et al.\ 1998).

\begin{figure}
\centerline{\psfig{file=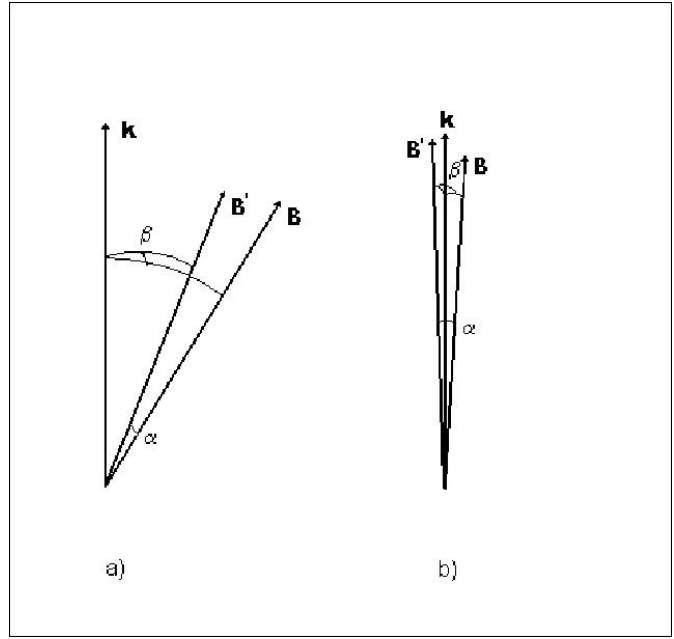,width=8.8cm,clip=} }
\caption{The turn of the $\bf k\times B$ plane caused by a variation of the
magnetic field along the ray path. If the angle  between the wave 
direction and the magnetic field is large ({\it a}), a  turn of 
$\bf B$ by a small angle
$\alpha$ result in a  turn of the  $\bf k\times B$ plane by a small angle
$\beta$. If the angle between $\bf k$ and $\bf B$ is small ({\it b}), a small
turn of $\bf B$  may result in a large turn of  the  $\bf k\times B$ plane.
\label{image}}
\end{figure}

\subsection{Orthogonal polarization modes and depolarization}
The plasma in the pulsar magnetosphere is birefringent; the orthogonally 
polarized ordinary and extraordinary waves propagates independently producing
the abrupt orthogonal transitions in polarization position angle that are
commonly observed in studies of individual pulse polarization (Manchester 
et al.\ 1975; Stinebring et al.\ 1984a,b; Gil \& Lyne 1995; Gangadhara 1997;
McKinnon,  Stinebring 1998, 2000). 
The observed depolarization of 
pulsar average profiles with increasing radio frequency (Manchester et al.\ 
1973; Morris et al.\ 1981; Xilouris et al.\ 1996; 
von Hoensbroech et al.\ 1998; Kramer et al.\ 1999) may be attributed to
merging of the ordinary and extraordinary beams at high frequency. 
McKinnon (1997)  assumed,
following Barnard \& Arons (1986), that pulsars emit originally in
both radiation modes and all the emission comes from a narrow range of 
heights above the stellar surface. The extraordinary mode propagates along 
straight ray
paths, while the ordinary waves may be deflected from the original 
direction by refraction. At  low frequencies, refraction
separates the two modes resulting in a net polarization. At high frequencies 
(above the local plasma frequency), 
refraction is negligible and both modes merge producing a non-polarized beam.

It is still unclear if it is possible to generate the extraordinary waves deep
inside the magnetosphere where the magnetic field is effectively infinite.
In this case, the extraordinary wave, whose electric vector is 
orthogonal to the background magnetic field, does not interact with the plasma
and cannot be emitted, at least in the homogeneous magnetic field. In 
principle, interaction is possible in a curved magnetic field, especially if
one takes into account that the characteristic frequency of the curvature
emission in pulsars is of the order of the emitted radio frequency. 
Unfortunately, nonlinear wave processes in a curved, super strong magnetic field
have not been considered yet. 

The extraordinary wave may appear in the outgoing radiation due to propagation
effects even if initially only ordinary waves were emitted. Because the 
plasma production process in pulsars may be unsteady,
the plasma flow in the open field line tube may be strongly inhomogeneous. One can
naturally assume that the plasma is gathered in  clouds separated by 
regions of the very low density such that the condition
$\Delta nkL\ll 1$ is satisfied in the inter-cloud space and the waves propagate
there preserving the polarization position angle. Now let us
consider an ordinary wave emitted within some cloud. Because the frequency of
this wave is of the order of the local plasma frequency,  refraction deviates
the ray from the initial direction such that the ray may come to another 
plasma cloud where the orientation of the local magnetic field is different
from that in the original cloud. Because adiabatic walking ceases in the
inter-cloud space, the ray enters the second cloud not in a normal mode
corresponding to the local orientation of the magnetic field. Therefore 
the wave is split into two independently propagating normal waves. This process
may produce orthogonal polarization modes and may result in depolarization
of the outgoing radiation. 

The depolarization mechanism outlined above works if the refraction angle is
comparable with the angular width of the open field line tube. Taking into 
account that the refraction angle is determined by the transverse density
gradient, which decreases with the increasing of the tube width, one can
state a general trend:  the narrower the open field line tube 
the less degree of polarization. Then the radius-to-frequency mapping
implies that the high frequency radiation, which is emitted at lower
heights where the tube is narrow, should be less polarized,  what is
indeed observed. Another observational finding may be considered as a partial
case of the above general trend. Namely because the width of the open 
field line tube decreases with the increasing pulsar period, depolarization
should be more pronounced in slow pulsars. The observed high frequency
polarization actually decreases with the pulsar period (Morris et al.\ 1981;
von Hoensbroech et al.\ 1998).

Recently Petrova (2001b) considered another mechanism for the
conversion of ordinary waves into extraordinary ones. Due to refraction, 
the ray may for a short while become nearly aligned with the local magnetic 
field. In this degenerate situation, refraction indexes of the two waves
become equal and the conversion takes place. Calculations were made in the
assumption that the  plasma distribution in the 
open field line tube is  axisymmetric, then the refracted ray remains in the
plane of the magnetic field line. In the general case, co-planarity
between the refracted beam and the field line fails and the ray may never 
become aligned with the magnetic field. Nevertheless
this effect may reveal itself at some phases of the pulsar period.

\subsection{Refraction}
Refraction was already invoked in the above discussion of the depolarization.
Of course the shape of  pulses should be also strongly affected 
by refraction. The observed pulse widths typically broaden with decreasing
frequency however at high frequencies ($>1$ GHz) the beam width vs.\ frequency 
curve commonly flattens (Sieber et al.\ 1975; Rankin 1983b; Thorsett 1991).
Barnard \& Arons (1986) attributed this behavior to the refraction. They    
assumed that all the pulsar emission comes from the same height; then  
the frequency dependence of the beam width  follows the frequency
dependence of the refraction angle: at high frequencies refraction is
negligibly small and the beam width is constant whereas at low frequencies
the beam widens with decreasing frequency due to increasingly strong
refraction. 

In the model advocated here, radiation is emitted at about the local
plasma frequency, which implies the standard radius-to-frequency mapping
$\nu\propto r^{-3/2}$. In this case the refraction index, which depends on
the ratio $\nu/\nu_p$, is the same for any ray at the emission point. However
the emission height and the width of the open field line tube increases
with decreasing frequency therefore the relative contribution of the 
refraction into the beam width decreases. Lyubarskii \& Petrova (1978)
considered the impact of refraction upon the beam width assuming the standard
radius-to-frequency mapping and  the axisymmetrical  
plasma distribution in the tube. The ray deviation occurs mainly on account
of the density gradient across the tube. If the plasma density decreases
towards the edge of the tube, the rays deviate outwards. Then at high 
frequencies the refraction angle exceeds the angular width of the open field tube
and the beam width becomes independent of the frequency whereas
at low frequencies the beam widens together with the open field line tube.
If the plasma density increases towards the edge of the tube (the density is
minimal at the magnetic axis), the ray deviate towards the magnetic axis and,
provided the tube is narrow enough, may even reflect from the opposite side
of the tube. As a result a trough arises at the beam width vs.\ frequency
curve, which resembles the so called ``absorption feature'' observed in some
pulsars (Rankin 1983b).

Assuming a hollow cone distribution of the plasma in the open field line tube
(the plasma density is small at the axis, increases outwards and then decreases
towards the edge) one can obtain pulse profiles with three  components 
(Petrova \& Lyubarskii 2000; Petrova 2000) resembling those observed in 
real pulsars. The rays deviating towards the axis form the core beam
whereas the rays deviating outwards form the cone beam. Petrova (2001a)
demonstrated that the refraction may also account for the unusual apparent
structure of the emission region inferred from the observations of the 
 interstellar scintillations (Wolszczan \& Cordes 1987; Gupta et al.\ 1999). 

The refraction may affect also the observed pulsar spectrum. Sieber (1997)
demonstrated that many properties of the observed spectra may be attributed
to the geometry of the beam. Petrova (2002) studied these effects in case the
geometry of the beam is determined by refraction.

Of course the assumption about the hollow cone plasma distribution is rather
restrictive. The plasma density at the axis should be small if the magnetic
field is strictly dipolar down the stellar surface; then the plasma production
rate is small at the magnetic axis. However it is naturally to believe that higher
order multipoles contribute to the field at the surface (see however Arons 
1993)
and then the plasma distribution across the tube is not so regular.
Nevertheless one can anticipate that many qualitative features of the ray 
behavior remain in more general configurations. Let us assume, for example,
that the instant plasma distribution in the tube is clumpy because the plasma
production process is unsteady. Refraction in such a clumpy medium  results
in some scatter in the ray directions, which provides a minimal beam width
even if the tube becomes very narrow. So saturation of the beam width vs.\
frequency curve at high frequencies (Sieber et al.\ 1975; Rankin 1983b; 
Thorsett 1991)  seems to be a rather general feature.

\section{Conclusion}
It was demonstrated above that a large variety of the observed properties
of the pulsar radio emission may be explained self-consistently, at least
at the qualitative level, assuming that the emission is generated by the
two-stream instability. Of course the theory falls far short of being 
complete. Evidences for a large transverse size
of the emission region inferred from the observations of the interstellar
scintillations (Smirnova et al.\ 1996; Gwinn et al., 1997, 2000; 
Hirano \& Gwinn 2001; Smirnova
\& Shishov 2001) make a challenge. A possible solution 
is the scattering of the beam on the plasma
inhomogeneities in the upper pulsar magnetosphere  
(Lyutikov \&  Parikh 2000; Lyutikov 2001). Another difficult problem is posed
by the recent finding that the Vela pulsar emits
in the extraordinary polarization mode (Lai et al., 2001). Plasma processes in
an effectively infinite magnetic field are believed to emit predominantly in the
ordinary mode, at least in the straight magnetic field. Possible emission of the
extraordinary mode in the curved magnetic field should be considered. 

\begin{acknowledgements}
 The support by the Heraeus foundation is gratefully acknowledged.
\end{acknowledgements}


\clearpage

\end{document}